\documentclass[%
 reprint,
 amsmath,amssymb,
 prf,onecolumn,
]{revtex4-2}
\usepackage[utf8]{inputenc}
\usepackage{physics}
\usepackage{graphicx}% Include figure files
\usepackage{bm}% bold math
\usepackage[colorlinks=true, allcolors=blue]{hyperref}
\newcommand\We{\textrm{We}}
\newcommand\D{d}

\newcommand\R{R_0}

\newcommand\force{\mathcal{T}}

\begin{document}
\title{Bubble breakup probability in turbulent flows}

\author{Ali\'enor Rivi\`ere$^{1,2}$}
\email{alienor.riviere@epfl.ch}
\author{St\'ephane Perrard$^2$}
\affiliation{$^1$LFMI, School of Engineering, EPFL, Lausanne, 1000, Switzerland}
\affiliation{$^2$PMMH, CNRS, ESPCI Paris, Universit\'e PSL, Sorbonne Universit\'e, Universit\'e de Paris, 75005, Paris, France}

\begin{abstract}
Bubbles drive gas and chemical transfers in various industrial and geophysical contexts, in which flows are typically turbulent. As gas and chemical transfers are bubble size dependent, their quantification requires a prediction of bubble breakup.
The most common idea, introduced by Kolmogorov and Hinze, is to consider a sharp limit between breaking and non breaking bubbles, given by $\We_c\approx 1$, where the Weber number $\We$ is the ratio between inertial and capillary forces at the bubble scale.
Yet, due to the inherent stochasticity of the flow every bubble might in reality break. In this work, we use a stochastic linear model previously developed to infer the breakup probability of bubbles in turbulence as function of both We and the residence time. This allows us to introduce a definition of the critical Weber number accounting for the time spent by bubbles within a turbulent region. We show that bubble breakup is a memoryless process, whose breakup rate varies exponentially with $\We^{-1}$. The linear model successfully reproduces experimental breakup rates from the literature. We show that the stochastic nature of bubble breakup is central when the residence time of bubbles is smaller than ten correlation times of turbulence at the bubble scale: the transition between breaking and non breaking bubbles is smooth and most bubbles can break. For large residence times, the original vision of Kolmogorov and Hinze is recovered. 
\end{abstract}

\maketitle

\section{Introduction}
Bubbles play a crucial role in mass transport across interfaces. By increasing the surface of exchange they lead chemical and gas transfers in various industrial processes, as homogenizers~\citep{haakansson2019}, bubble column reactors~\citep{kantarci2005,han2007,risso2018}, and geophysical situations, such as rivers, waterfalls~\citep{beaulieu2012,demars2013} and oceans~\citep{melville1996,woolf2007,deike2022}. 
Since gas exchanges are bubble size dependent, predicting breakups in turbulent flows is central to understand the bubble size distribution and then quantify the transfers. 
In inertial flows, bubble fate is controlled by the ratio between inertial and capillary forces, namely, the Weber number, $\We = \rho U^2 \D/\gamma$ where $\rho$ is the fluid density, $\D$, the bubble equivalent diameter, $U$ a typical velocity and $\gamma$ the surface tension.
Low Weber number bubbles are stable, while they break above a critical Weber number $\We_c$ which depends on the flow geometry~\citep{muller2008,chu2019,rodgar2023,riviere2023}.
It is worth mentioning that in the case of dense turbulent emulsions \citet{crialesi2023} defined the critical Weber number as the one for which the net energy flux from capillarity is zero.
In turbulence, according to Kolmogorov and Hinze theory~\citep{kolmogorov1949bubble,hinze1955} bubbles are mainly deformed by eddies of their size so that $U$ can be estimated as the average velocity increment at the bubble scale $\langle \delta u(\D)^2\rangle$. 
In homogeneous and isotropic turbulence (HIT), for size lying within the inertial range, the velocity increment relates to the energy dissipation rate $\epsilon$~\citep{kolmogorov1941}, which quantifies the energy transfers across scales, through $\langle \delta u(\D)^2\rangle = 2 (\epsilon \D)^{2/3} $.
The Weber number then reads $\We = 2\rho \epsilon^{2/3}\D^{5/3}/\gamma$.
Note that this definition relies on average quantities. It does not account for the inherent fluctuations of the flow. As a consequence, since any bubble might encounter a large enough pressure fluctuations which breaks it, in turbulence $\We_c$ is only defined in a statistical sense~\citep{vela2022,ni2023}.
In addition, in practical situations, flows are inhomogeneous (bubble columns) or unsteady (breaking waves) or both, and the time spent by a bubble within an homogeneous turbulent region, called the residence time, affects the critical Weber number $\We_c$. To the best of our knowledge, the residence time has not yet been taken into account in theoretical models.

In this article, we quantify the statistics of bubble lifetime using a linear stochastic model for bubble deformations, whose parameters were inferred using direct numerical simulations (DNS)~\cite{riviere2024}. Our approach is based on the reminiscent idea that non linear effects may be negligible until the critical deformation for breakup is reached~\citep{risso1998}, as introduced by several authors for bubbles~\citep{risso1998,galinat2007,lalanne2019,masuk2021simultaneous} and drops~\citep{roa2023}.
However, all previous studies used models with parameters derived in the quiescent case, in particular for the damping rate~\citep{rayleigh1879,lamb1932}, which turns out to be one order of magnitude larger in the presence of a turbulent background flow. 
It is worth mentioning that \citet{ravelet2011} experimentally observed and reported a damping rate much higher than in a quiescent flow for bubble rising in turbulence. They attributed this additional damping to the bubble wake.

This article is organized as follows. We first introduce the stochastic linear model. Then, running our reduced dynamics, we measure the probability for a bubble to break as a function of We and the residence time. We deduce the associated breakup rate and compare it to experimental datasets.
We eventually introduce a definition of the critical Weber number accounting for the residence time and discuss practical applications in inhomogeneous and unsteady flows.

\section{Linear stochastic oscillator}

\begin{figure}
\centering
\includegraphics{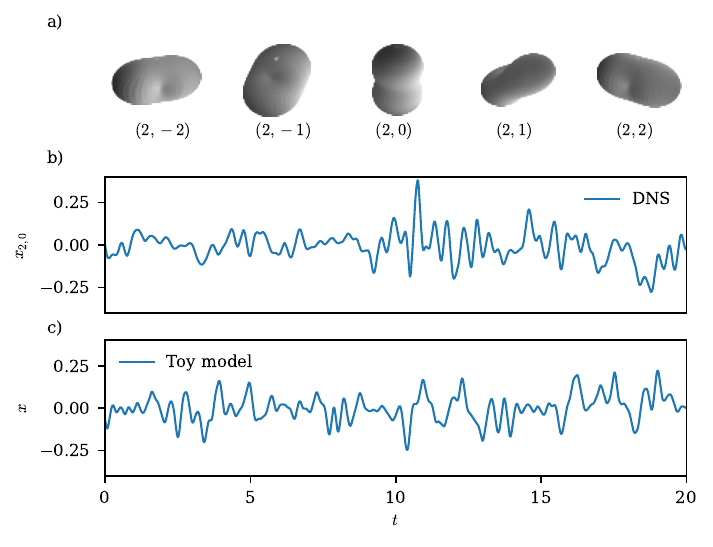}
\caption{a)~Shape of deformation of the five modes $\ell =2$. b)~Typical temporal evolution of a mode $\ell=2$, at $\We = 0.71$ measured in a DNS~\cite{riviere2024}. c)~Typical temporal evolution given by the linear stochastic model defined by equation \eqref{eq:model}, at the same Weber number $\We = 0.71$.}
\label{fig:evolution}
\end{figure}

For small amplitude deformations, the bubble surface can be described by the local radius $R(\theta, \phi)$ with $\theta$ the colatitude and $\phi$ the longitude. This parametrization holds as long as the local radius is mono-valued in $(\theta, \phi)$, corresponding to limited bubble deformations, and it will fail close to breakup. In a co-moving frame of reference, the local bubble radius $R$ decomposed onto the real spherical harmonics base $Y_\ell^ m(\theta, \phi)$ reads,
\begin{equation}
    R(\theta, \phi) = \R\big[1 + \sum_{\ell=2}^\infty \sum_{m=-\ell}^\ell x_{\ell, m}(t)Y_\ell^m(\theta, \phi)\big]
\end{equation}
where $\R$ is the volume-equivalent bubble radius. Since the frame of reference moves with the bubble, the three harmonics $(\ell=1)$, encoding bubble translation, are null at all time.
Note that the spherical harmonics decomposition is not invariant under frame rotation: rotating the reference frame redistributes the weight associated to a mode $(\ell, m)$ on all the modes $(\ell, m^\prime)_{m^\prime \in [-\ell, \ell]}$.
In turbulence, bubble shape continuously reorients due to random fluctuations and torques.
Since we do not prescribe any special frame orientation relative to the bubble shape, we cannot distinguish the dynamics of $x_{\ell, m}$ from the ones of $x_{\ell, m^\prime}$ with $m^\prime \neq m$. Hence the $2\ell +1 $ modes $\ell $ will have equivalent dynamics.
Each mode amplitude $x_{\ell, m}$ evolves over time under the stochastic forcing imposed by the surrounding turbulent flow.
In turbulence, bubbles are known to mainly deform and break following oblate-prolate oscillations, which correspond to the five modes $(\ell = 2)$~\citep{risso1998,perrard2021} illustrated on Figure~\ref{fig:evolution}a.
In the following, we will therefore focus on the dynamics of these five modes.
We denote by $x$ the amplitude of one of them. As an illustration, we show on figure~\ref{fig:evolution}b the temporal evolution of one mode $\ell=2$ at $\We = 0.71$ measured in a DNS following the procedure described in \citep{riviere2024}.

\begin{figure}
\centering
\includegraphics{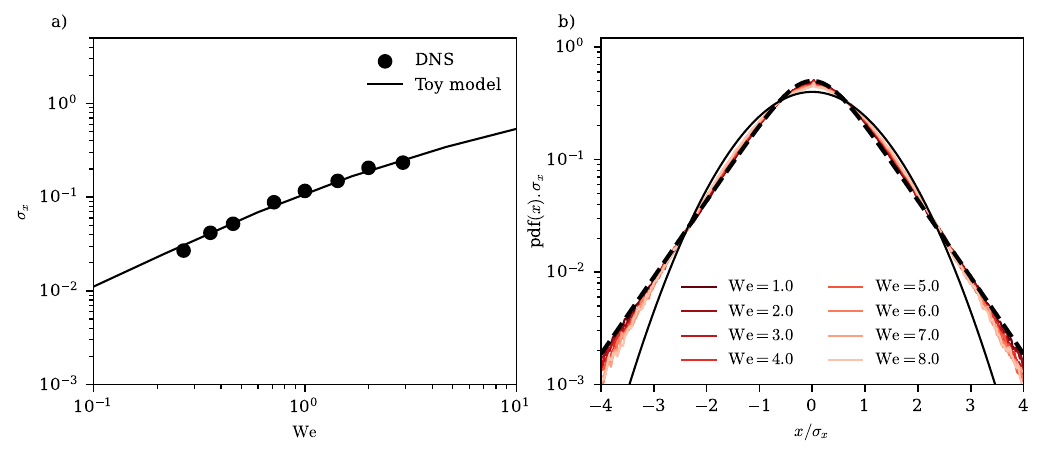}
\caption{a) Standard deviation of modes $\ell=2$ as a function of the Weber number from the DNS~\cite{riviere2024} (black dots) and given by the model~\eqref{eq:model} (solid line). b) Normalized distributions of the mode amplitude obtained by running the model~\eqref{eq:model} for various Weber numbers. The black dotted line is the hyperbolic secant distribution, which is the probability distribution of the stochastic forcing $\force$. For comparison, the black solid line shows a Gaussian distribution.}
\label{fig:comp}
\end{figure}

We recently showed~\cite{riviere2024} that each mode $\ell = 2$ dynamics can be described in a statistical sense by a linear oscillator randomly forced by the surrounding flow,
\begin{equation}
    \ddot{x} + \Lambda \dot{x} + \Omega^2 x = \force(t),
    \label{eq:model}
\end{equation}
where lengths are in unit of the bubble equivalent radius $\R$, and times are in unit of the eddy turnover time at the bubble scale $t_c=\epsilon^{-1/3}d^{2/3}$, which encodes the correlation time of turbulent structures of bubble's size. The values of $\Lambda$ and $\Omega$ as well as the statistical properties of $\force$ were previously inferred from a DNS~\cite{riviere2024}. We found that the dimensionless damping coefficient $\Lambda=12$ is independent of the Weber number. At $\Re(d)= \sqrt{2}\epsilon^{1/3}d^{4/3}/\nu = 124$, with $\nu$ the liquid kinematic viscosity, this value is about 7 times higher than the viscous dissipation in a quiescent fluid. Indeed, turbulence induces an effective turbulent damping, as observed for drop shape oscillations in the presence of an internal turbulent flow~\citep{berry2005,xiao2021}. As the bubble lies within the inertial range, where the flow is scale invariant, we assume that $\Lambda$ also does not depend explicitly on the Reynolds number $\Re(d)$.
Note that for drop breakup, the study of \citep{vela2022} implies that the drop breakup rate is independent of the Reynolds number at the drop scale and hence the underlying deformation dynamics. The natural frequency remains unchanged compared to the quiescent value~\citep{rayleigh1879,lamb1932} so that $\Omega^2 = 192/\We $.
The forcing term $\force$ models the erratic forcing from turbulence by a time-correlated random variable independent of the bubble properties, and therefore of We. The statistically stationary forcing $\force$ is fully characterized by its probability distribution function (pdf), and its auto-correlation function. The pdf of $\force$ can be modelled by an hyperbolic secant distribution $1/(2\sigma_\force)/\cosh{(\pi x/(2\sigma_\force))}$ with a standard deviation  $\sigma_\force = 20$ independent of We. This pdf coincides with the pdf of the pressure mode $\ell =2$ integrated over a sphere of radius $\R$ in the absence of bubble~\cite{riviere2024}.
The auto-correlation function writes $\exp(-2\pi. 2^{2/3}t)(1 + 2\pi.2^{2/3}t)$, where the $2^{2/3}$ factor originates from the eddy turnover time at the scale $d/\ell$ with $\ell=2$. Note that the typical correlation time of the effective forcing is $0.2 t_c$, which is significantly shorter than the eddy turn-over time $t_c$. Eventually, equation~\ref{eq:model} models adequately the dynamics of one mode $\ell=2$ of amplitude $x$. 
We use the linear model to generate a large sample of mode dynamics. We first generate 5000 independant temporal sequences of $\force$ using the procedure described in the Supplementary Material I. Each temporal sequence lasts 50$t_c$ and has a temporal resolution $\Delta t_1 = 0.01t_c$. This choice ensures to have 10 points per correlation time of the forcing. Starting with an initial condition $(x(0),\dot x(0))$, we integrate over small time steps $\Delta t_2/\Delta t_1 = 0.1$, the exact solution of Eq.~\ref{eq:model} for a $\force$ piecewise constant. We test the sensitivity to the initial conditions, and found that they are forgotten on a time scale of order $0.2 t_c$ corresponding to the correlation time of the forcing, which is much shorter than a typical break-up time (see Supplementary Material II). In the following, we therefore use the 5000 temporal evolutions of $x$ of length $50 t_c$, with initial conditions  $x=0, \dot x = 0$, corresponding to a initially spherical bubble at rest.
A typical dynamics of the reduced model~\eqref{eq:model} is shown on Figure~\ref{fig:evolution}c for $\We = 0.71$. The evolution of the standard deviation $\sigma_x$ with the Weber number is captured by the linear model, as shown in figure~\ref{fig:comp}a. The pdf of $x$ shown in figure~\ref{fig:comp}b is also described by an hyperbolic secant distribution, as expected from a linear model. 

\section{A memoryless process}

\begin{figure}
\centering
\includegraphics{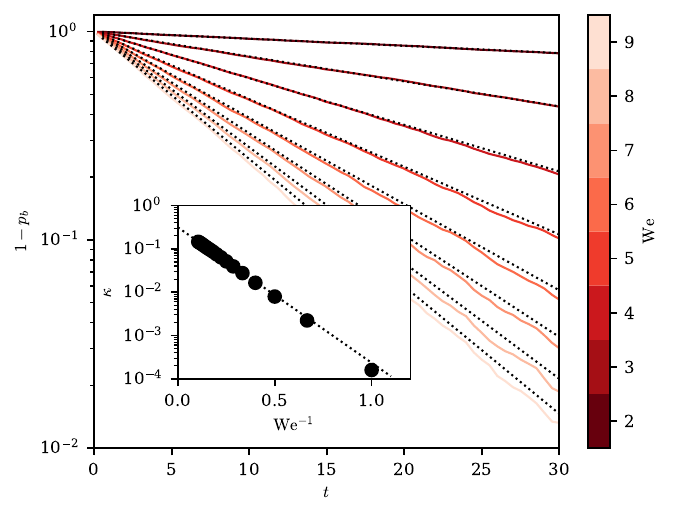}
\caption{Survival probability associated to one mode, as a function of time for several We (solid color lines) in the model~\eqref{eq:model}. Black dotted lines are exponential fits for each Weber number. Inset plot: associated breakup rate $\kappa$ as a function of We. The black dotted line is an exponential fit, see equation~\eqref{eq:kappawe}. }
\label{fig:proba}
\end{figure}

In order to model breakup, we introduce a critical linear deformation at which a bubble breaks.
We choose $x_c = 0.74$ for consistency with our previous work~\citep{perrard2021}. 
Accordingly, we define the breakup time of one oscillator $x$ as the first time when the amplitude reaches $x_c$.
The cumulative probability, $p_b(t)$, is the probability that one mode reaches $x_c$ before an observation time $t$.
Figure~\ref{fig:proba} shows the survival probability associated to one mode $\ell =2 $, $1-p_b(t)$, as a function of the observation time, for several Weber numbers.
The survival probability decays exponentially in time, as expected for a memoryless process, with a constant breakup rate $\kappa$ associated to each mode of deformation. For a breakup time much larger than the correlation time of both the velocity fluctuations at the bubble scale and the bubble deformations, the dynamics can indeed be modeled as a succession of independent events. Eventually, the breakup probability associated to one mode $\ell = 2$ reads $p_b(t) = 1 - \exp[-\kappa(\We) t]$.

We now connect the breakup associated to one mode $\ell=2$ with bubble breakup. We assume that a bubble breaks when one of the five modes $\ell=2$ reaches the critical deformation.
These modes, in the linear limit, are uncoupled and follow the dynamics given by equation~\eqref{eq:model}.
As a consequence, the probability that a bubble breaks before $t$ is given by $P_b(t) = 1-(1-p_b(t))^5$ and can be written as
\begin{equation}
    P_b(t) = 1 - \exp[-5\kappa(\We) t],
    \label{eq:proba}
\end{equation}
which extends to all Weber numbers the previous experimental observation of memoryless bubble breakups at large Weber number by \citet{ravelet2011}. This type of memoryless breakup process was also observed numerically for drop breakups~\citep{vela2022}.
We find that the bubble breakup rate $\kappa_b$ is five times larger than that of one single oscillator: $\kappa_b = 5 \kappa$.
A similar argument has been used by \citet{brouzet2021} to model fiber breakup in turbulence.
In their case, since a fiber is described as a collection of $N$ elementary rigid elements, the number of breakup modes can be approximated by the number $N$ of elements.

\begin{figure}
\centering
\includegraphics{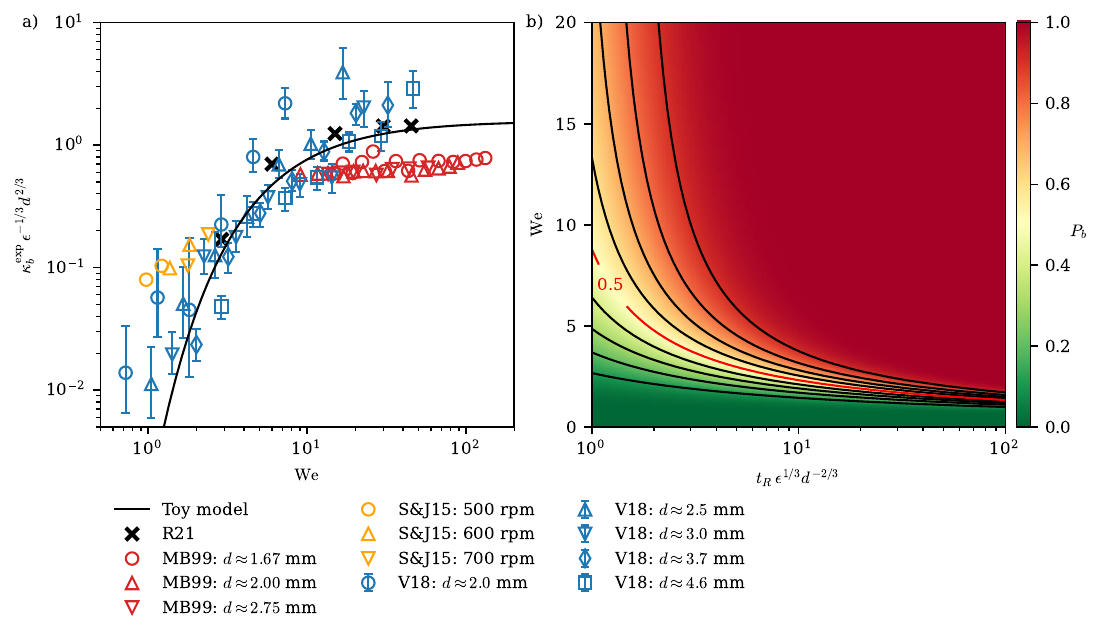}
\caption{a)~Comparison between our predicted breakup rate and several experimental datasets: \citet{martinez1999a} (MB99), \citet{solsvik2015} (S\&V15)\citet{vejravzka2018} (V18), \citet{riviere2021} (R21). b)~Probability for a bubble to break depending on its We and the residence time $t_R$ normalized by the eddy turnover time. Black lines are isoprobabilities separated by 0.1. The red line corresponds to a breakup probability of $1/2$.}
\label{fig:compdata}
\end{figure}

The breakup rate associated with one mode $\ell=2$, $\kappa$, varies exponentially with $\We^{-1}$ (see inset figure~\ref{fig:proba}).
It follows that the bubble breakup rate $\kappa_b$ reads
\begin{equation}
\kappa_b = 5 \kappa = \kappa^\infty \exp[\frac{\We^t}{ \We}] 
\label{eq:kappawe}
\end{equation}
%\kappa = 0.314 old definition
with $\kappa^\infty = 1.57$ and $\We^t =7.20 $, two numerical factors obtained by a least-square fit.
$\kappa^\infty$ is the breakup rate in the limit of infinitely large $\We$.
Its inverse, $1/\kappa^\infty=0.64 $, of the order of the eddy turnover time, represents the associated bubble lifetime in the absence of surface tension and is controlled by the typical advection time of the two bubble sides toward each other \citep{perrard2021}.
The transitional Weber number, $\We^t$ sets the transition between fast breaking and unlikely slow breakups.
This law, which suggests a mechanism of random activation process, was first proposed by \citet{coulaloglou1977} in the context of emulsions, based on the idea of drop-eddy collisions.
Note that when bubble size lies within the inertial range of the turbulent cascade,  bubble dynamics are inertial and we expect equation~\eqref{eq:kappawe} to hold for all moderate Reynolds number, as similarly shown recently for drops~\citep{vela2022}.
Figure~\ref{fig:compdata}a compares $\kappa_b$ with the breakup rates measured in various experimental conditions~\citet{martinez1999a,solsvik2015,vejravzka2018}, $\kappa_b^\textrm{exp}$ in s$^{-1}$, made dimensionless by the eddy turnover time $t_c$ at the bubble scale, together with the DNS data from \citet{riviere2021}. The experimental data points were originally gathered together in~\citet{zhong2024}, and are reproduced here with courtesy of Pr Rui Ni. Note that, experimental breakups rates are estimated using different expressions, which induces an additional scattering of the data~\citep{haakansson2020}.
Even though the data set were obtained in different experimental facilities with various turbulence intensities corresponding to different bubble Reynolds number, our linear model follows the experimental points. This result confirms the intuition of \citet{risso1998} that bubble lifetime distribution can be predicted using a linear model. Note that the present model only describe independent break-up events. The lifetime distribution of rapid correlated events such as filament breaking~\cite{riviere2022} do not fall under the scope of model~\eqref{eq:proba}.

\section{Breakup probability and critical Weber number}
In real flows, homogeneous turbulent regions have a finite extent in both space and time. As a consequence, bubbles remain only a finite time, the residence time, $t_R$ in s, within the turbulent region with a probability to break before $t_R$ of
\begin{equation}
    P_b(t_R, \We) = 1 - \exp[\frac{- \kappa^\infty t_R}{ \epsilon^{-1/3}d^{2/3}} \exp(-\frac{\We^t}{\We})].
    \label{unifying}
\end{equation}
This equation predicts, that any bubble whose size lies within the inertial range, can break, provided the residence time is long enough.
However, in practice, residence time are finite and the breakup of bubbles at sufficiently small We is unlikely to occur. The curve $P_b(t_R, \We_c) = 0.5$ defines the critical Weber number. Figure~\ref{fig:compdata}b shows the probability to break as a function of both We and the residence time in unit of the eddy turnover time.
For short residence times, $P_b$ gently varies with We. As a consequence, there is a wide range of We for which a bubble may break ($0.1<P_b<0.9$). The probabilistic nature of the breakup process smooths the transition around $P_b=0.5$.
At long residence time compared to the eddy turnover time ($t_R>40 t_c$), the limit between breaking and non breaking bubbles sharpens. We conclude that the probabilistic nature of bubble breakup in turbulence is essential at short resident times while the deterministic vision of Kolmogorov and Hinze is recovered for large resident times ($t>O(10 t_c$).

Even though other mechanisms might dominate over turbulent fluctuations in some practical situations, such as shear and vorticity~\citep{joseph1998,muller2008,chu2019,riviere2023}, it is tantalizing to compare the results from our linear stochastic model to unsteady and inhomogeneous flows. In what follows, we discuss three different situations - plunging jets, oceanic bubbles and industrial bubble columns - which correspond to three different regions of the diagram of figure~\ref{fig:compdata}b.
For example, for a typical plunging jet, with speed 5~m.s$^{-1}$~\citep{kiger2012} at the interface, the bubble jet depth is about $50$~mm, leading to a residence time of $0.01$~s.
The energy dissipation rate can be estimated using the velocity fluctuations $u^\prime$, which is typically 0.1\% of the entrance velocity~\citep{kiger2012}, and the integral lengthscale which is the nozzle radius $D \sim 10$~mm, giving $\epsilon = (u^\prime)^3/D\sim 10$~m$^2$s$^{-3}$.
As a consequence, the residence time is of the order of two eddy turnover times. At this residence time merely all bubbles can break (see figure~\ref{fig:compdata}b). We expect the coexistence of large bubbles with much smaller ones, leading to a broad bubble size distribution, which is indeed the case below plunging jets and waterfalls.

In the oceans, underneath breaking waves, turbulence is sustained for approximately 1/3 of the wave period, which is of the order of one to few seconds, with a turbulent intensity of 1~m$^2$s$^{-3}$.
For centimetric to millimetric bubbles the residence time ranges from $6$ to $30$ eddy turnover time, corresponding to a critical Weber number between 3 and 5, consistent with the experimental measurements~\citep{deane2002}.
In industrial bubble columns, bubbles rise with a velocity of about $0.1$~m.s$^{-1}$ on meter columns, inducing a weak turbulence of around $0.1$~m$^2$s$^{-3}$~\citep{kantarci2005}.
The residence time for millimetric bubbles is then of the order of 1000 eddy turnover time, where the breaking transition is extremely sharp. In this regime, all bubbles will have $\We<\We_c\approx 1$.

In this article, we use a stochastic linear model for bubble deformations to quantify the bubble break-up probability. We show that the breakup statistics can be adequately modeled by considering a critical deformation for breakup for any of the five modes of deformation. This result validates the idea of Risso and Fabre \cite{risso1998} that bubble lifetime can be predicted from a linear deformation model. We provide an explicit expression to compute the break-up probability as a function of the bubble Weber number and the resident time. We introduce a definition of the critical Weber number which accounts for the resident time, unifying the probabilistic vision of bubble breakup with the original idea of Kolmogorov and Hinze. Provided that turbulent fluctuations dominate over other breaking mechanisms, the model can be applied to practical situations with finite resident time such as unsteady and inhomogeneous flows.

\begin{acknowledgements}
We thank Rui Ni and Shijie Zhong for sharing the data used to generate figure 4.
We also thank Luc Deike, François P\'etr\'elis, S\'ebastien Gom\'e, Louis-Pierre Chaintron and Emmanuel Villermaux for fruitful discussions.
This work was performed using HPC resources from GENCI-IDRIS (Grant 2023-AD012B14107). This work was also granted access to the HPC resources of MesoPSL financed
by the Region Ile de France and the project Equip@Meso (reference
ANR-10-EQPX-29-01) of the programme Investissements d'Avenir supervised
by the Agence Nationale pour la Recherche. S. Perrard aknowledges financial supports from PSL junior fellow grant and from the National Research Agency (reference ANR-23-CE30-0043-03). 
\end{acknowledgements}

\bibliography{biblio}

\end{document}